# An Explanation of the Differences in Diffusivity of the Components of the Metallic Glass $Pd_{43}Cu_{27}Ni_{10}P_{20}$

K. L. Ngai[1*], and S. Capaccioli[1,2]

[1]*Dipartimento di Fisica, Università di Pisa, Largo B. Pontecorvo 3, I-56127, Pisa, Italy*

[2]*CNR-IPCF, Institute for Physical and Chemical Processes, Largo B. Pontecorvo 3, I-56127, Pisa, Italy*

**Abstract**

Bartsch *et al*. [A. Bartsch, K. Rätzke, A. Meyer, and F. Faupel, Phys. Rev. Lett. **104**, 195901 (2010)] reported measurements of the diffusivities of different components of the multi-component bulk metallic glass $Pd_{43}Cu_{27}Ni_{10}P_{20}$. The diffusion of the largest Pd and the smallest P were found to be drastically different. The Stokes-Einstein relation breaks down when considering the P constituent atom, while the relation is obeyed by the Pd atom over 14 orders of magnitude of change in Pd diffusivity. This difference in behavior of Pd and P poses a problem challenging for explanation. With the assist of a recent finding in metallic glasses that the β-relaxation and the diffusion of the smallest component are closely related processes by Yu *et al*. [H. B. Yu, K. Samwer, Y. Wu, and W. H. Wang, Phys. Rev. Lett. **109**, 095508 (2012)], we use the Coupling Model (CM) to explain the observed difference between P and Pd quantitatively. The same model also explains the correlation between property of the β-relaxation with fragility found in the family of $(Ce_xLa_{1-x})_{68}Al_{10}Cu_{20}Co_2$ with $0 \leq x \leq 1$.

*Author to whom correspondence should be addressed:
ngai@df.unipi.it

## I. Introduction

Bulk metallic glasses (BMG) are constituted of metal or metalloid atoms interacting with pair potential. Distinctly different in chemical composition and binding from the common molecular glass-formers, the study of the dynamic properties of BMG offers opportunity in advancing the understanding of the liquid state and transition to the glassy state. Particularly relevant is the identification of the dynamic properties of BMG that are commonly found in molecular glass-formers and considered basic. If identified, any such dynamic property would receive additional support for its fundamental importance and universal manifestation in glass-formers in general. The secondary or β-relaxation of BMG is such an example. It is either resolved in the isochronal mechanical loss spectra [1-31-3], or presented as an excess wing on the high frequency (low temperature) side of the isothermal (isochronal) spectra [4,5]. There is only one secondary relaxation present in BMG. In contrast, more than one secondary relaxation are found in many small molecular or polymeric glass-formers, and their properties enable to separate them into two different families [6]. The secondary relaxations involving internal or intra-molecular degree of freedom have no connection with the structural α-relaxation. On the other hand, the secondary relaxation involving rotation/translation of the entire molecule, or the repeat unit in the case of polymer, have strong connections to the structural α-relaxation in dynamic properties [6-9]. Naturally intermolecular secondary relaxations of this kind have fundamental significance, and to distinguish them from the unimportant secondary relaxations involving motion of part of the molecule, they are called the Johari-Goldstein (JG) β-relaxation [10].

Since there is only one secondary relaxation in BMG, it is likely the analogue of JG β-relaxation in molecular glass-formers. This possibility seems real in view of several recent findings by experiments of the correlation between the β-relaxation of BMG with properties some of which are directly related to the structural α-relaxation. Here we give several examples.

(1) The crystallization of BMG at temperatures below the glass transition temperature under the ultrasonic vibrations is caused by accumulation of atomic jumps associated with the *β*-relaxation being resonant with the ultrasonic strains [11].

(2) The activation of shear transformation zones (STZs) and β-relaxations in metallic glasses are directly related, with the activation energy of the β-relaxation nearly the same as the potential-energy barriers of STZs [12].

(3) The β-relaxation of $La_{68.5}Ni_{16}Al_{14}Co_{1.5}$ is closely correlated with the activation of the structural units of plastic deformations and global plasticity, and the brittle to ductile transition and the β-relaxations follow similar time-temperature dependence [13].

(4) The dynamical mechanical properties of a series of BMG, $(Ce_xLa_{1-x})_{68}Al_{10}Cu_{20}Co_2$ with $0 \leq x \leq 1$, show that the properties of the β-relaxation are closely correlated with the fragility of the supercooled liquids [14].

(5) Most recently, Yu et al. [15] demonstrated for BMG in the glassy state that the diffusion motion of the smallest constituent atom occurs within the temperature and time regimes where the β-relaxation is activated, and there is good agreement between the activation energies of the two processes.

These experimental facts and particularly (5) enable not only better understanding of the nature of the β-relaxation, but also allow us to make connection with the diffusivities data of the radiotracers, $^{103}Pd$, $^{32}P$, $^{57}Co$, and $^{51}Cr$, in a $Pd_{43}Cu_{27}Ni_{10}P_{20}$ melt, where the smallest and largest atoms are P and Pd respectively [16]. The diffusivities were measured by Bartsch et al. [16] over a wide temperature range. While the diffusivities of all components are the same at temperatures above 710 K, decoupling in diffusivity was observed between the slower Pd and of the smaller components starting at about 710 K. Below 710 K, the decoupling increases with decreasing temperatures to reach 4 orders of magnitude at the glass transition temperature $T_g$. The first task of this paper is to quantitatively address the decoupling using the result from Yu et al. [15], i.e. property (5), that the diffusion motion of the smallest constituting atom of BMG and the β-relaxation are related, and they have the same activation energies in the glassy state.

Remarkably the Stokes-Einstein (SE) relation between the viscosity of $Pd_{43}Cu_{27}Ni_{10}P_{20}$ and the diffusivity of Pd holds over the whole temperature range investigated encompassing more than 14 orders of magnitude in the change of the diffusivity, which is in stark contrast to the breakdown of the SE relation found in van der Waal molecular glass-formers including *ortho*-terphenyl, trinaphthal benzene [~~16~~17-22], sucrose benzoate [23], and indomethacin [24] from experiments. This observation of Pd is a surprise because the structural relaxation of BMG is collective as shown by isotope mass dependence experiments and simulations [25], and hence is dynamically heterogenerous like that found in other glass-formerssuch as the van der Waals glass-formers [26], and colloidal suspensions [27]. Spatially heterogeneous dynamics had been used to explain the breakdown of SE relation simply as a result of the difference in how self diffusion and structural relaxation or viscosity are averaged over the distribution of time scales [28-31]. Supported by various direct and indirect experimental evidences, there is no doubt that the structural relaxation is dynamically heterogeneous in most if not all glass-formers. However, the way it was used before to explain the difference in temperature dependence between diffusion and structural relaxation or viscosity is now recognized as inconsistent with experiments [20-24,30-37] and simulations [37-39]. The recent Perspective paper published by Ediger and Harrowell in the same journal [40] has made this amply clear by the statement:

"Initially, it was suggested that the difference in temperature dependence between diffusion and structural relaxation, for example, arose as a result of the difference in how the respective observables averaged over the distribution of time scales. This view is now seen as inconsistent with experiments". Notwithstanding, computer simulations of liquids at short times have given hope that the difference in the temperature dependence of the translational and rotational diffusion constants can be similarly understood by a growing length scale associated with heterogeneous dynamics, and dynamic heterogeneity may in some other way can still be used to explain the breakdown of SE relation, the authors of the Perspective [40] are careful to point out that computer simulations are restricted to short relaxation times and, hence the results are limited to higher temperatures, and it is possible that other physics governs the dynamics at lower temperatures where the breakdown of SE relation is most prominent.

Thus, at the present time, it is not clear how to use the spatially heterogeneous dynamics in $Pd_{43}Cu_{27}Ni_{10}P_{20}$ to explain the experimental results of Bartsch et al. [16], although its existence in glass-formers is certain and it occupies an important role in understanding dynamics properties, a belief the Coupling Model that we use here holds and shares with the research community at large. Given this situation, it is challenging to rationalize why the SE relation between viscosity and diffusivity is strictly observed by Pd but not by P in the multi-component BMG, and not in the single-component molecular glass formers while all are dynamically heterogeneous.. This is the second task undertaken in this paper. As the third and final task, we explain the experimental finding of a relation between β-relaxation and fragility of the structural α-relaxation in LaCe-based BMG [14]. All explanations of the experimentally observed effects considered in this paper will be given from the Coupling Model, a comprehensive review of its theoretical basis and various applications can be found in Ref.[7].

**II. Decoupling between Pd and the smaller components in diffusivity**

Bartsch et al. [16] measured the radiotracer diffusivities of Pd, P, Cr, and Co in $Pd_{43}Cu_{27}Ni_{10}P_{20}$ melt by using as radiotracers the isotopes, $^{103}Pd$, $^{32}P$, $^{51}Cr$, and $^{57}Co$. The measurements were carried out over a wide range of temperatures above the caloric glass transition temperature $T_g$ of 582 K that were obtained at a heating rate of 20 K/min. Their data are combined with diffusivity data of P, Ni, and Pd from others. Taking viscosity data over the same temperature range from the literature, possible decoupling between viscosity $\eta$ and the diffusion coefficient $D$ of the components were examined. To compare with viscosity data, the diffusion data are converted to diffusion viscosity, $\eta_D$, by using the Stokes-Einstein (SE) relation, $D=k_BT/6\pi\eta_Dr$, where $k_B$ is the Boltzmann constant, and $r$ the particle radius. If the SE relation is obeyed by a component, its $\eta_D$ will coincide with the measured viscosity $\eta$. This is the case for Pd, and the SE relation was found to hold over more than 14 orders of magnitude change of $D$ in the entire range of temperature above $T_g$ investigated. The diffusion viscosity $\eta_D$ of the smaller component P as well as the radiotracers $^{57}Co$, and $^{51}Cr$ are in agreement with the measured viscosity $\eta$ only at temperature above $T_c$=710 K. Below $T_c$, $\eta_D$ falls below $\eta$ and the difference increases monotonically with decreasing temperature to reach 4 orders of magnitude on approaching $T_g$=582 K. Thus the smaller P, Co and Cr strongly violate the SE relation.

More than 12 years ago, an alternative explanation of the breakdown of SE relation and Debye-Stokes-Einstein (DSE) relation was proposed [41], which is based on the Coupling Model (CM) of relaxation and diffusion in many-body interacting systems. The CM explanation was offered as an alternative explanation, but it was eclipsed by the popular and intuitively more appealing explanation that the difference in temperature dependence between diffusion and structural relaxation originates from the difference in how the respective observables averaged over the distribution of time scales of spatially heterogeneous dynamics [29-31]. However the premise of this popular explanation has been found to be contradicted by the temperature independence of the time or frequency dispersion of the structural relaxation of tri-naphthal benzene (TNB) [20,32,33], ortho-terphenyl (OTP) [21,22], sucrose benzoate [23], and indomethacin [24,36]. Thus the way to explain breakdown of SE relation from spatially heterogeneous dynamics is inconsistent with experiments, as concluded in a recent review ]40], although there is no doubt that the structural relaxation is dynamically heterogeneous. On the other hand, the CM explanation published in 1999 [41] continues to hold in view of these experimental findings as demonstrated in Ref.[34]. Actually, decoupling has been found between two observables not involving translational diffusion [41-45], and also in glassy ionic conductors [46-49]. Hence breakdown of SE relation is a special case of a more general phenomenon, which has been rationalized by the CM based on different observables can have different coupling parameters [41-49].

The CM is based on the many-body relaxation/diffusion in interacting systems caused by inter-particle or intermolecular interaction/coupling, which obviously include the glass-forming liquids. Since many-body relaxation is spatially random and is not homogeneous, the structural relaxation of the CM is dynamically heterogeneous. This was made clear first in 1990 in Ref.[50], and reiterated in Refs.[41] and [34]. This property is seldom emphasized or used in applications of the CM, and hence it might have led others including Sillescu [50] to think that the CM is homogeneous. But this mistaken view of the CM was rectified in a follow-up paper published by Sillescu with others [52], where they cited the 1990 CM paper [50].

As mentioned in the above, the dynamic heterogeneous property was not utilized in all applications of the CM. Instead other features of the CM were used to address the various experimental findings including the decoupling of two observables in general, and in particular the breakdown of the SE relation as follows. The alternative CM explanation is based on the, thesis that the effect of the many-body dynamics is weighed differently on different observables. This is modeled by different observables, $\mu$, have different coupling parameters, $n_\mu$. The parameter $n_\mu$ quantifies the slowing down of the structural α-relaxation when probed in terms of the observable $\mu$ by the many-body effects or cooperativity originating from intermolecular interaction . It is the complement of the fractional exponent, $(1-n_\mu)$ of the Kohlrausch stretched exponential correlation function for the observable $\mu$,

$$\langle \mu(0)\mu(t)\rangle / \langle \mu^2(0)\rangle = \exp[-(t/\tau_\mu)^{1-n_\mu}], \tag{1}$$

The measured relaxation time, $\tau_\mu$, is given by the CM equations [7,41,42,47,49],

$$\tau_\mu(T) = [t_c^{-n_\mu} \tau_{0\mu}(T)]^{1/(1-n_\mu)}, \tag{2}$$

where $\tau_{0\mu}$ is the primitive relaxation times of dynamic variables or observable $\mu$, and $t_c$ is the temperature insensitive time of crossover from primitive relaxation to many-body relaxation [7]. It can be seen from Eq.(2), anomalous properties of $\tau_\mu$ are generated from properties of $\tau_{0\mu}$ by raising the it to the power of $1/(1-n_\mu)$. Uninfluenced by the cooperative many-body relaxation dynamics, all properties of $\tau_{0\mu}$ are normal. These include that $\tau_{0\mu}$ is governed by the same friction coefficient for all observables $\mu$, and hence all $\tau_{0\mu}$ have the same temperature dependence, or the same primitive activation energy $E_0$, if the temperature dependence is Arrhenius.

Eq.(2) together with different $n_\mu$ for different $\mu$ immediately lead to differences between the $\tau_\mu$'s and their temperature dependencies, and hence decoupling. A larger $n_\mu$ for the observable $\mu$ will bestow stronger temperature dependence for the relaxation time. If all observables are in the Arrhenius regime where $\tau_\mu$ has activation energy $E_\mu$, from Eq.(2) we have the relation,

$$\frac{E_\mu}{E_0} = (1 - n_\mu), \tag{3}$$

with different $E_\mu$ follows from different $n_\mu$.

Applied to the multi-component $Pd_{43}Cu_{27}Ni_{10}P_{20}$ melt, the predominant metallic bonds have no fixed directions and hence no permanent relations among the components. Thus, the diffusion of each component has its own coupling parameter for the correlation function represented by Eq.(1) with µ now used to label the components. The coupling parameters $n_D$ of diffusion of the smaller components are smaller than that of the largest Pd majority component because smaller atom has lesser inter-atomic constraints. From the CM Eq.(2) it follows that the diffusion coefficients of the smaller components have weaker $T$-dependence than that of Pd. Hence this difference between the $T$-dependence of the diffusivity of Pd and the diffusivities of smaller components explains qualitatively their decoupling which increases with decreasing temperature, and reaches more than 4 orders of magnitude at the glass transition temperature $T_g =$ 582 K. The qualitative explanation given above has limitations because the values of $n_D$ of P and Pd are not provided by the data of diffusivities. This less than favorable situation is like that encountered in first using the alternative CM explanation of the breakdown of the SE and DSE relations of single component molecular glass-formers [7,34,41], where $n_D$ of self-diffusion or probe diffusion is not provided by the measurement of diffusivity. Only in a rare case of ionic liquids [41], $n_D$ was known together with the coupling parameter of the structural relaxation $n_\alpha$ or viscosity $n_\eta$, and the qualitative explanation becomes rigorous. In the present case of $Pd_{43}Cu_{27}Ni_{10}P_{20}$, we are fortunate to have the results from the recent study by Yu et al. [15], and from which we can deduce the coupling parameter $n_D$ of P. On the other hand, we can obtain $n_\alpha$ or $n_\eta$ from the mechanical relaxation data of a closely related BGM, $Pd_{40}Ni_{10}Cu_{30}P_{20}$, [2]. Consequently, a quantitative explanation of the decoupling of diffusivity between Pd and P becomes possible as will be carried out below.

We use the CM Eqs.(1) and (2) in conjunction with the findings of Yu et al. [15] to explain quantitatively the decoupling between Pd and the smaller P, Co, and Cr in diffusivity. The key result from Yu et al. is the diffusion motion of the smallest constituting atom occurring within the temperature and time regimes where the β-relaxation are activated, and the two processes have almost the same activation energy. This important and general result shows that diffusion of the smallest component, P in the present case, and the β-relaxation are closely related. Experiments carried out in many different kinds of glass-formers [7] including metallic glasses [5] have shown the relaxation time, $\tau_\beta$, of the β-relaxation can be identified with the primitive relaxation time $\tau_0$. On combining these two properties, the temperature dependence of the diffusion coefficient of P, $D_P$, can be identified with that of the primitive relaxation time $\tau_{0\mu}$ appearing in Eq.(2). In the Arrhenius regime, the activation energy $E_P$ of $D_P$ is the primitive activation energy $E_0$ common to all observables μ, including the viscosity $\eta$ and the structural relaxation time $\tau_\alpha$. Substituting $E_P$ for $E_0$ in Eq.(3), we have

$$\frac{E_\alpha}{E_P} = (1 - n_\alpha), \quad \frac{E_\eta}{E_P} = (1 - n_\eta) \qquad (4)$$

The activation energies $E_\alpha$ and $E_\eta$ of the α-relaxation and viscosity are the same, which is a consequence of the Maxwell relation, $\eta = G_\infty <\tau_\alpha>$. It follows from $E_\alpha=E_\eta$ and Eq.(4) that $n_\alpha$ and $n_\eta$ are equal.

Later on we shall make comparison of diffusivities data of P and Pd with mechanical relaxation data of the BMG taken at low frequencies and temperatures not far above $T_g$. Therefore we consider the temperature dependencies of $D_P$ and $D_{Pd}$, the diffusion coefficient of Pd measured at the lower end of the experimental temperature range, where the apparent activation energies of $D_P$ and $D_{Pd}$ can be determined. The lines drawn by Bartsch et al. in their Fig.1 are used to estimate the diffusivity activation energies $E_P$ and $E_{Pd}$, and their ratio $E_{Pd}/E_P$ in a limited temperature range, 590<$T$<625 K, not far above $T_g$. The ratio we obtain is given by

$$E_{Pd}/E_P \approx 1.79. \tag{5}$$

Since Pd obeys the SE relation, the viscosity $\eta$ and the structural relaxation time $\tau_\alpha$ from shear mechanical relaxation, $\tau_\alpha$, have the same activation energy as $E_{Pd}$ for diffusivity in this limited temperature range, i.e.,

$$E_\eta = E_\alpha = E_{Pd} \tag{6}$$

Here we have ignored the minor variation of the factor, $T$, in the SE relation. By combining Eqs.(4), (5), and (6), and together with $n_\eta=n_\alpha$, we have the result,

$$E_{Pd}/E_P \approx 1/(1-n_\eta) = 1/(1-n_\alpha). \tag{7}$$

In Eq.(7), $n_\alpha$ is the coupling parameter of the shear mechanical structural α-relaxation, which can be obtained by fitting the Fourier transform of the Kohlrausch function to the frequency dependence of the measured shear mechanical relaxation spectrum of the BMG to be discussed next.

On substituting the ratio $E_{Pd}/E_P$ from Eq.(5) into Eq.(7), the value of $(1-n_\alpha)$ is determined by

$$(1 - n_\alpha) = 0.56 \qquad (8)$$

This is the value predicted for the stretch exponent of the Kohlrausch function of the structural α-relaxation in Eq.(1), where the dynamic variable μ therein is the shear modulus. If the value of the fractional exponent given by Eq.(8) is correct, it should be the same as the stretch exponent of the Kohlrausch function fitting the shear modulus data of $Pd_{43}Cu_{27}Ni_{10}P_{20}$ melt in the same temperature range, $590<T<625$ K, Such measurements have not been made on $Pd_{43}Cu_{27}Ni_{10}P_{20}$, but can be checked by experiments carried out in the future. At this time Young's modulus has been measured in the melt of another BMG, $Pd_{40}Ni_{10}Cu_{30}P_{20}$, with composition nearly the same as $Pd_{43}Cu_{27}Ni_{10}P_{20}$, at temperatures in the range of $573 \leq T \leq 593$ K near $T_g=593$ K [2]. The Kohlrausch function was used after Fourier transform to fit the isothermal Young's modulus data of $Pd_{40}Ni_{10}Cu_{30}P_{20}$ by Zhao et al. [2]. They reported the value of 0.57 for the stretch exponent, $\beta_{KWW} \equiv (1-n_\alpha)$, in the Kohlrausch function used in the fit. This value is close to the value of 0.56 given by Eq.(8). Thus we conclude that the CM can explain quantitatively the decoupling of diffusivity between Pd and P observed by Bartsch et al.[16].

## III. Relation between $\beta$ −relaxation and fragility in LaCe-based BMG

Observation of the changes of characteristics of the structural α-relaxation systematically on varying the chemical composition of BMG can be helpful to understand the mechanism of glass transition. Moreover, any relation of the β-relaxation to the structural α-relaxation that can be established in BMG would be enlightening to understand the connection that β-relaxation has to various properties of BMG mentioned in the Introduction. These objectives are not easy to realize in the study of BMG in contrast to molecular glass-formers because characterization of dynamics of BMG is limited to dynamic mechanical measurements over limited frequency range. Notwithstanding, a recent dynamic mechanical study by Yu et al. [14] of the series of BMG belonging to the same family, $(Ce_xLa_{1-x})_{68}Al_{10}Cu_{20}Co_2$ where $0 \leq x \leq 1$, managed to make some progress in this direction. From the structural α-relaxation time $\tau_\alpha$ determined over a range of temperatures in the supercooled liquid state, the fragility index, $m = d\log\tau_\alpha/d(T_g/T)$ evaluated at $T_g/T=1$, was obtained for the series of BMG. The isochronal mechanical loss $E''$ spectra of the series at 1 Hz were compared after scaling temperature by $T_g$ and normalizing the α-loss peaks, located at $T/T_g=1$, to have the same height. In this plot of normalized $E''$ versus $T/T_g$, they found the following correlation between fragility of the supercooled liquid and a property of the β-relaxation. More fragile BMG with larger $m$ has more intense β-relaxation peak and is further separated from the α-loss peak (*i.e.*, located at a lower value of $T/T_g$ in the isochronal mechanical loss spectrum). The correlation resembles some of the correlations between $\tau_\beta$ and $\tau_\alpha$ found in molecular glass-formers [6-9,54]. In particular is the correlation of the intensity of β-relaxation and its separation from the α-relaxation given by $\log(\tau_\alpha/\tau_\beta)$ with the non-exponentiality of the α-relaxation or $n_\alpha$ appearing in the stretch exponent $(1-n_\alpha)$ of the Kohlrausch correlation function. If restricted to glass-formers of the same family, non-exponentiality or $n_\alpha$ of the α-relaxation usually correlates with fragility $m$ [7,55,56]. Applying this to the family of $(Ce_xLa_{1-x})_{68}Al_{10}Cu_{20}Co_2$, the correlation found between the β-relaxation properties and the fragility of the α-relaxation by Yu et al. can be restated as correlation between $\log(\tau_\alpha/\tau_\beta)$ and $n_\alpha$. As demonstrated before for molecular glass-formers [6-9,33], this correlation is expected from the CM Eq.(2) together with the fact that $\tau_0$ is approximately the same as $\tau_\beta$ [6-9,54]. This is shown explicitly by recasting Eq.(2) for μ≡α into the form,

$$\log\left(\frac{\tau_\alpha}{\tau_0}\right) = (\tau_0/t_c)^{n/(1-n)} \approx (\tau_\beta/t_c)^{n/(1-n)} \tag{9}$$

Since the value of $\tau_\beta$ from dynamic mechanical measurement of Yu et al. is much longer than $t_c \approx 1$ ps, and the exponent $n/(1-n)$ in Eq.(9) is a monotonically increasing function of $n$, the correlation between $\log(\tau_\alpha/\tau_\beta)$ and $n_\alpha$ follows as a consequence of Eq.(9).

**IV. Conclusion**

Bulk metallic glasses (BMG) and particularly those with multi-components, such as $Pd_{43}Cu_{27}Ni_{10}P_{20}$, are formed by collection of atomic particles held together by metallic bonds. They differ greatly in chemical bonding and structure from molecular and colloidal glass-formers, and one may expect they exhibit dynamic properties related to glass transition that are different from single-component molecular glass-formers. Measurements of radiotracer diffusivities of all components in a $Pd_{43}Cu_{27}Ni_{10}P_{20}$ melt over a broad temperature range down to near $T_g$ reported recently by Bartsch et al. indeed present new phenomena that are challenging to explain. These include decoupling between the diffusivity of Pd and of the smaller components, which increases with decreasing temperature to reach more than 4 orders of magnitude at the glass transition temperature $T_g$. Unexpectedly, the Stokes-Einstein relation holds for Pd in the whole range investigated encompassing more than 14 orders of magnitude change in diffusivity. We demonstrated that the Coupling Model (CM) can explain the data of Bartsch et al. not only qualitatively, but also quantitatively with the assist of the finding that diffusion of the smallest component in BMG and secondary β-relaxation are closely related.

The β-relaxation of BMG have been found to bear connection to important properties including crystallization, activation of shear transformation zones, and brittle-ductile transition. These connections may originate from the β-relaxation serving as the precursor of the structural α-relaxation. This possibility is strengthened by the observation a correlation between the intensity and relaxation time of the β-relaxation and fragility of the α-relaxation in the family of BMG, $(Ce_xLa_{1-x})_{68}Al_{10}Cu_{20}Co_2$. We show this correlation is like that found in molecular and polymeric glass-formers, and the correlation is expected from the CM.